\documentclass{aastex62}
\usepackage{amsmath}
\usepackage[sort&compress]{natbib}
\usepackage{soul}

\shorttitle{C-Complex Space Weathering Trends}
\shortauthors{Thomas et al.}

\begin{document}

\title{Space Weathering Within C-Complex Main Belt Asteroid Families}

\correspondingauthor{Cristina A. Thomas}
\email{cristina.thomas@nau.edu}

\author[0000-0003-3091-5757]{Cristina A. Thomas}
\affil{Northern Arizona University,
Department of Astronomy and Planetary Science\\
PO Box 6010,
Flagstaff, AZ 86011 USA}

\author{David E. Trilling}
\affiliation{Northern Arizona University,
Department of Astronomy and Planetary Science\\
PO Box 6010,
Flagstaff, AZ 86011 USA}

\author{Andrew S. Rivkin}
\affiliation{Johns Hopkins University Applied Physics Laboratory}

\author{Tyler Linder}
\affiliation{University of North Dakota}



\begin{abstract}

Using data from the Sloan Digital Sky Survey (SDSS) Moving Object Catalog, we study color as a function of size for C-complex families in the
Main Asteroid Belt to improve our understanding of space weathering of carbonaceous materials. We find two distinct spectral slope trends: Hygiea-type and Themis-type.
The Hygiea-type families exhibit a reduction in spectral slope with increasing object size until a minimum slope value is reached and the trend reverses with increasing slope with increasing object size. The Themis family shows an increase in spectral slope with increasing object size until a maximum slope is reached and the spectral slope begins to decrease slightly or plateaus for the largest objects. Most families studied show the Hygiea-type trend. The processes responsible for these distinct changes in spectral slope affect several different taxonomic classes within the C-complex and appear to act quickly to alter the spectral slopes of the family members. 

\end{abstract}

\keywords{}


\section{Introduction} \label{sec:intro}

Large spectral and spectrophotometric datasets enable powerful studies of the compositions and surface properties of small body populations in our Solar System. In particular, the Sloan Digital Sky Survey (SDSS) Moving Object Catalog is an excellent resource that contains photometric observations of over 100,000 unique known moving objects. Previous studies have used the SDSS Moving Object Catalog photometry to investigate the distribution of taxonomic types across the Main Belt \citep{carvano2010,demeo2013}, the colors of the Main Belt families \citep{parker2008}, and space weathering trends both between families \citep{nesvorny2005SDSS} and within certain families \citep{thomas2012}. All of these works have demonstrated that the SDSS photometry can be reliably used to distinguish between taxonomic classes and determine spectral slopes. The ability of this dataset to determine spectral slopes for large numbers of objects is particularly important to the study of spectral trends associated with space weathering. 

Spectral changes due to space weathering have been extensively studied for S-complex asteroids \citep[e.g.][]{binzel2004,thomas2011,thomas2012,gaffey2010}, ordinary chondrite meteorites \citep[e.g.,][]{strazzulla2005,moroz1996}, the Hayabusa samples \citep[e.g.,][]{noguchi2011}, and other silicate-rich materials \citep[e.g.,][]{sasaki2001,pieters2000,brunetto2005,loeffler2009}. Many investigations have concluded that silicate-rich materials display similar signs of space weathering including increased spectral slope, decreased band depth, and decreased albedo. In the past decades our understanding of how space weathering affects the physical properties of S-complex material has continuously improved, while our knowledge regarding the C-complex materials has lagged behind. 

\cite{nesvorny2005SDSS} examined space weathering for the S and C-complex families across the Main Belt by investigating the change in principal color components compared to the age of the family. Their analysis supported the conclusion that S-complex materials show an increased slope with increasing age and was the first to suggest that the spectral slopes of C-complex materials show a blueing trend with increased age. 
Complementary studies of carbonaceous chondrites have shown that laboratory space weathering simulation experiments on different meteorite types have resulted in conflicting observed spectral slope changes. The laboratory observations were summarized in \cite{lantz2017}: irradiation of CV and CO chondrites results in redder spectral slopes, while CI and CM chondrite spectra become bluer. 
All of these published C-complex trends match a key expectation of space weathering induced spectral changes that the trend should continue with time up to the point when saturation is reached and no further change occurs \citep[e.g.,][]{pieters2000}. 

We examine space weathering trends within Main Belt asteroid families using the Sloan Digital Sky Survey (SDSS) Moving Object Catalog. By considering each family independently, we remove variation in composition from our analysis of changing spectral properties. The assumption that composition is consistent within each family is common, but imperfect \citep[e.g.,][]{masiero2015}. For most families included in this study, we do not have evidence to suggest compositional variation within the family. The case of the partially differentiated Themis parent body is discussed further in \S\ref{sec:disc}. 
We investigate trends of spectral slope with respect to asteroid size for all known C-complex asteroid families with sufficient data. We assume that smaller objects have younger surfaces on average since the expected collisional lifetime \citep{obrien2005,deelia2007}, the timescale for regolith refresh via seismic shaking \citep{richardson2005}, and the potential to experience YORP spin-up and failure \citep{rubincam2000,graves2018} are dependent on the size of the asteroid.

\section{Data \& Analysis} \label{sec:analysis}

The Sloan Digital Sky Survey (SDSS) Moving Object Catalog (MOC, \cite{ivezic2002asteroids}) consists of near simultaneous observations in five photometric filters ($u$, $g$, $r$, $i$, and $z$) with effective center wavelengths of 3551, 4686, 6166, 7480 and 8932 $\AA$, respectively \citep[e.g.,][]{ivezic2001,juric2002}. The use of the $g$, $r$, $i$, and $z$ filters provides sufficient wavelength coverage to determine likely taxonomic classifications \citep[e.g.,][]{carvano2010,demeo2013} and study spectral slope trends \citep[e.g.,][]{thomas2012,graves2018}.
We use the fourth and final data release of the SDSS MOC which contains photometry of over 470,000 moving objects including over 100,000 unique known objects. Our previous work \citep{thomas2012} used the older third data release due to non-photometric data in the fourth release. We have modified our analysis procedure to account for the non-photometric data. 

We investigated each C-complex asteroid family included in the \citet{nesvorny2015families} version 3.0 catalog. The Nesvorny family lists include the calculated $C_j$ parameter for each asteroid to identify potential dynamical interlopers in the family according to their V-shape criterion \citep{nesvorny2015identification}. Objects with $\lvert C_j\rvert > 1$ are suspected to be dynamical interlopers and are removed from the sample. This criterion was defined in \cite{nesvorny2015identification} and its usage is supported by analysis of the Eos family in \cite{vokrouhlicky2006}. Once these potential interlopers have been removed, we identify the family asteroids within the SDSS MOC. The Nesvorny family lists include asteroid numbers, while the SDSS MOC include columns for asteroid number and preliminary designation or name. Additionally, the last update to the SDSS MOC predates the Nesvorny catalogs by several years and many objects in the MOC were numbered in the intervening years. To properly identify any given family member within the SDSS dataset, we modified the {\em check observability} python script which uses the {\em callhorizons}\footnote{https://github.com/mommermi/callhorizons} module to query JPL Horizons. For each asteroid number, our modified routine returns the name (when available) and preliminary designation. To guarantee that each object observation is only returned once, we search the MOC for all asteroid names and preliminary designations within each family. 

\setcounter{table}{0}
\begin{table}[b]
\renewcommand{\thetable}{\arabic{table}}
\centering
\caption{C-Complex Families in the SDSS Moving Object Catalog} \label{tab:Cfamilies}
\begin{tabular}{clccccll}
\tablewidth{0pt}
\hline
\hline
\colhead{Asteroid} & \colhead{Asteroid} & \colhead{Nesvorny} &
\colhead{Number of} & \colhead{Bin} & \colhead{Spectral} & \colhead{Albedo} & \colhead{Family}\\
\colhead{Number} & \colhead{Name} & \colhead{ID} &
\colhead{Objects} & \colhead{Size} & \colhead{Type} & \colhead{} & \colhead{Age}\\
\hline
\multicolumn{8}{l}{Complete Families}\\
\hline
10 &	Hygiea &	601	& 326 & 65 &C	& 0.068\tablenotemark{a}	& 2 $\pm$1 Gy\tablenotemark{c}\\
24 &	Themis &	602	& 448 & 90 & C	 & 0.066\tablenotemark{a} & 2.3 Gy\tablenotemark{d} \\
128 & Nemesis & 504 & 52 & 10 & C	 & 0.071\tablenotemark{a}	 & 200 $\pm$100 My\tablenotemark{c}\\
145 & Adeona & 505	 & 244 & 50 & Ch & 0.059\tablenotemark{a} & 700 $\pm$500 My\tablenotemark{c}\\
668 & Dora & 512 & 131 & 26 &Ch & 0.06\tablenotemark{b} & 500 $\pm$200 My\tablenotemark{e}\\
\hline
\multicolumn{8}{l}{Incomplete Families}\\
\hline
31 &	Euphrosyne &	901 & 131 & 26  & Cb & 0.056\tablenotemark{a} & $<$1.5 Gy\tablenotemark{e}\\
163 & Erigone & 406	 & 130 & 26 & Ch & 0.051\tablenotemark{a} & 300 $\pm$100 My\tablenotemark{f}\\	
490 & Veritas &	609 & 99 & 20	& Ch	 & 0.066\tablenotemark{a} & 8.3 $\pm$0.5 My\tablenotemark{f}\\
1726	 & Hoffmeister & 519	 & 74 & 15 & C	& 0.06\tablenotemark{b} & 300 $\pm$ 200 My\tablenotemark{c}\\	
\hline
\multicolumn{8}{c}{$^{a}$\cite{masiero2013}, $^{b}$\cite{demeo2013}, $^{c}$\cite{nesvorny2005}, $^{d}$\cite{marzari1995},}\\
\multicolumn{8}{c}{$^{e}$\cite{broz2013},$^{f}$\cite{nesvorny2015identification}}
\label{table:allfam}
\end{tabular}
\end{table}

For each asteroid family, we remove unreliable SDSS MOC data according to various criteria. We remove all observations with apparent magnitudes greater than 22.0, 22.2, 22.2, 21.3, and 20.5 in any of the $u$, $g$, $r$, $i$, and $z$ filters, respectively. These are the limiting magnitudes for 95\% completeness \citep{ivezic2001,demeo2013}. To address the non-photometric nights included within the fourth release of the SDSS MOC, we remove observations with flags relevant to moving objects and good photometry: $edge$, $badsky$, $peaks too close$, $not checked$, $binned4$, $nodeblend$, $deblend degenerate$, $bad moving fit$, $too few good detections$, and $stationary$ \citep[as done in][]{demeo2013}. These flags address a variety of issues including if the object was too close to the edge of the frame, if the local sky was poorly determined, if the peak of the object was too close to another object to be deblended, or if the object was not detected to move. Additional information for each of these flags can be found in the documentation associated with the fourth SDSS MOC release\footnote{http://faculty.washington.edu/ivezic/sdssmoc/sdssmoc.html}. 

We exclude the $u$ filter observations due to the significantly higher errors on the observations and place a limit on photometric uncertainty on the remaining 4 filters of 0.06 magnitudes. This is a more restrictive uncertainty limit than \cite{thomas2012} and \cite{demeo2013}. The selection of 0.06 magnitudes as the uncertainty limit was driven by the spectral slope analysis of the C-complex families. Spectrophotometry of a C-complex object will have a small spectral slope with calculated slope errors that will be notably impacted by large photometric errors. We selected the photometric uncertainty limit to balance the need for accurate photometry while maintaining an adequate sample size in the family populations.  
We restrict our analysis to the 9 families that have SDSS photometry with a minimum of 50 unique objects. 

Table \ref{table:allfam} includes information on all of the C-complex families in our analysis including the parent asteroid number and name, the Nesvorny catalog number, the number of family objects in our analysis, the bin size for each family, the spectral type, albedo, and the age of the family. 
To ensure that any observed spectral trends are not the result of observational biases due to the incompleteness of a family, we examine the size frequency distribution of each of these 9 C-Complex families. A family is classified as complete in this analysis if the completeness limit for the family is at an H magnitude larger (diameter smaller) than any observed changes in our spectral slope trends (Section \ref{sec:results}). Five families are classified as complete. For those five families, we repeat our analysis to ensure that the inclusion of data from objects beyond the completeness limit does not impact our results. Details regarding how we determine completeness and our additional analysis of the five complete families are included in Appendix \ref{sec:append}.
We include all 9 families in our analysis because the incomplete families show spectral trends consistent with the complete ones, which suggests that the observed trends are properties of the families and are not a product of observational bias. 

To calculate the slope for each observation, we convert the SDSS magnitudes to reflectance values. We remove the known Sloan filter solar colors\footnote{http://classic.sdss.org/dr6/algorithms/sdssUBVRITransform.html} from each of the color indices: $g-r=0.44$, $r-i=0.11$, and $i-z=0.03$. The Sun-removed color indices are then converted into reflectance values, which are normalized to the $r$ band (6166 $\AA$). Each slope is calculated as a linear fit to the $g$, $r$, $i$, and $z$ reflectance values using the central wavelength of the filters. Slope values are given in \%/micron. 
The errors associated with these reflectance values are calculated using standard error propagation techniques. The error of the slope is determined by a Monte Carlo calculation. Each calculated reflectance value is modified with an offset equal to the propagated reflectance error multiplied by a random number found from a Gaussian distribution between -1 and 1. For each observation, this calculation is done 20,000 times and a slope is determined for each modified spectrum. The slope error is calculated as the standard deviation of the modified slopes generated by this process. To avoid weighting our analysis on objects that were observed many times by the SDSS, we average the slopes and propagate the slope errors for all objects with multiple observations. This is the same process as used in \cite{thomas2012}. In the Themis family, we rejected two likely interlopers within the family for having slopes many standard deviations higher than the surrounding sample.

We investigate overall trends in spectral slope with respect to object size by using a running box mean as was done in \cite{binzel2004}, \cite{thomas2011}, and \cite{thomas2012}. The bin size for each family is selected to be approximately 20\% of the total number of objects. We use a bin size of approximately half the fractional size of the bins chosen by Binzel et al.\ (50 for 145 objects) and Thomas et al. (35 for 90 objects; 150 for 402 objects). The narrower bin size enables the study of more structure within our slope trends. The error for the average slope value of each bin is the propagation of each individual slope error. 
We use Solar System absolute magntude, $H$, as our primary indicator of object size. We approximate object diameters using the H magnitudes provided by the SDSS MOC and the known average albedos of the families from \cite{masiero2013} if available. If no family average albedo is available, we use the average albedo by taxonomic complex for Main Belt objects given by \cite{demeo2013} (C-complex $p_{V}=0.06$). We use the taxonomic classifications given by \cite{masiero2015} for each family. 


\begin{figure}[t]
\epsscale{1.2}
\plotone{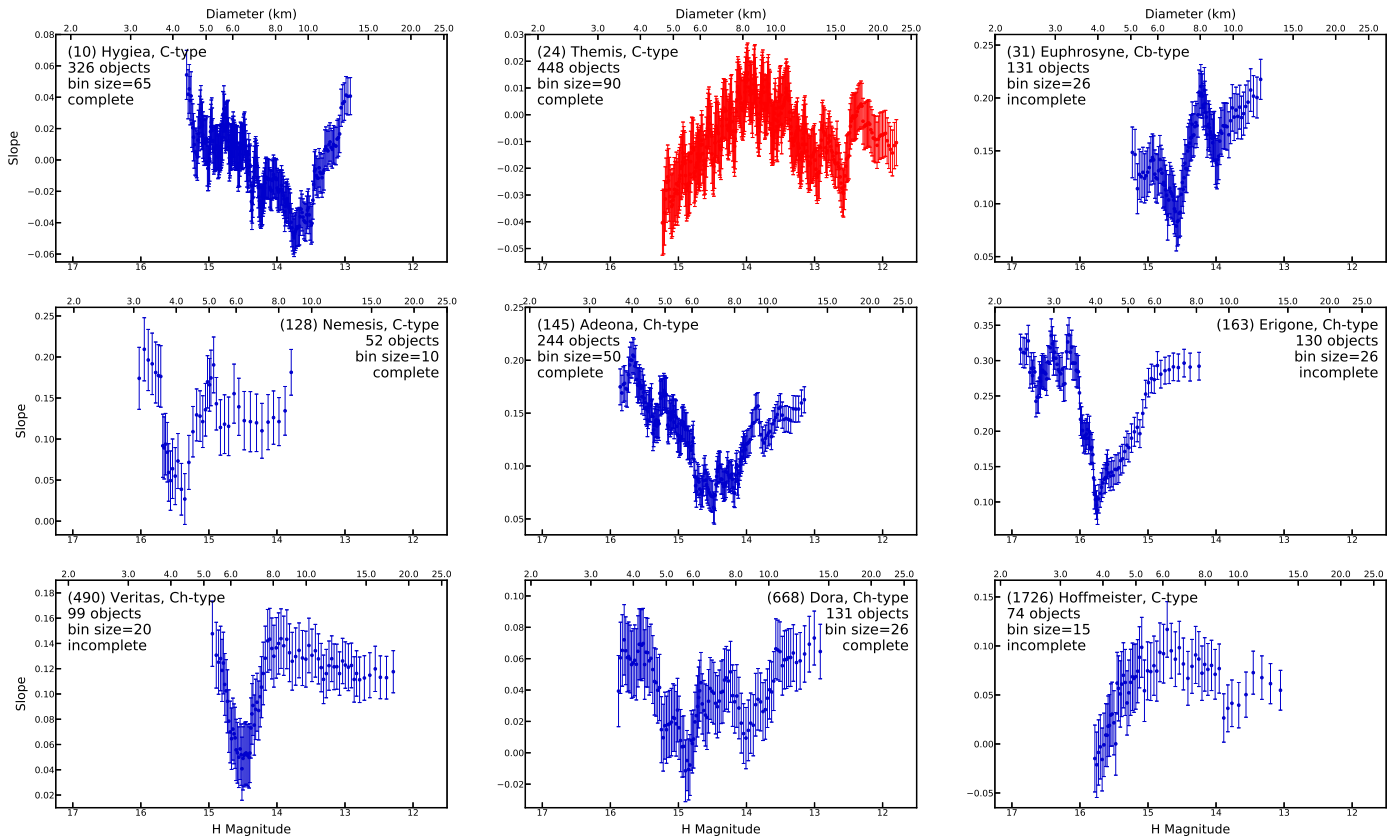}
\caption{Spectrophotometric slope versus H magnitude and approximate object diameter for the nine C-complex asteroid families that have at least 50 unique objects with a photometry magnitude error limit of 0.06 magnitudes. Each figure includes the family name, taxonomic type, number of objects in the final sample, the bin size for the running box mean calculation, and whether the family is complete or incomplete. The range of H magnitude is held constant for each family and is shown along the bottom axis for each figure. The top axis shows the approximate diameters for the corresponding H magnitudes using the average albedos for each family given in Table \ref{tab:Cfamilies}. The Hygiea-type families (blue) show a reduction in spectral slope with increasing object size until a minimum slope value is reached and the trend reverses with increasing slope with increasing object size. Section \ref{sec:results} and Figure \ref{fig:Hoff} contain further discussion regarding the classification of Hoffmeister as a Hygiea-type family. The Themis family (red) displays an increase in spectral slope with increasing object size until a maximum slope is reached and the spectral slope begins to decrease slightly for the largest objects. \label{fig:panelC}}
\end{figure}

\section{Results} \label{sec:results}

\begin{figure}[t]
\epsscale{0.7}
\plotone{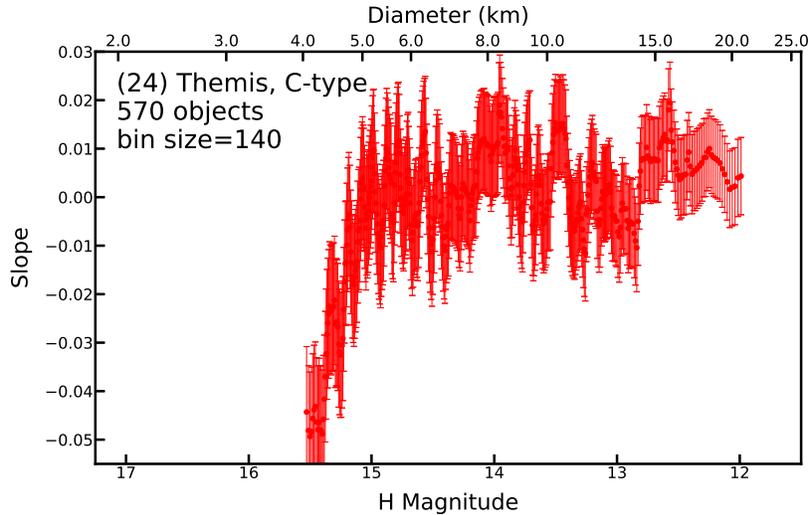}
\caption{Wtih a less selective error limit of 0.08 magnitudes, the Themis family shows a clear increase in spectral slope with increasing object size until the spectral slope plateaus at objects with diameters of approximately 5~km. \label{fig:Themis}}
\end{figure}

We see two distinct trends for the C-complex families: Hygiea-type and Themis-type. Figure 1 shows a panel of the 9 complete and incomplete C-complex families analyzed in this study. Each figure in the panel shows the binned spectrophotometric slopes versus H magnitude. Each figure also includes estimated diameter (calculated using the albedos in Table \ref{tab:Cfamilies}) along the top axis. For clarity, each family is labeled with their taxonomic type, the number of unique objects in the analysis, the bin size used, and whether the family is complete or incomplete. 
The Hygiea-type trend shows a clear reduction in spectral slope (blueing) with increasing object size until a minimum slope value is reached. The trend then reverses and the spectral slope increases (reddening) with increasing object size until the largest objects in the family are sampled (e.g., Hygiea family) or the spectral slope plateaus (e.g., Nemesis family). The spectral slopes tend to be approximately equal for the smallest and largest objects in the sample. We include the slope trends for the incomplete families in our analysis since all the families are consistent with the Hygiea-type trend. The Themis-type trend has a clear increase in spectral slope (reddening) with increasing object size until a maximum slope value is reached. At our nominal error limit (Figure \ref{fig:panelC}), the trend then reverses and the spectral slope decreases slightly. For a higher error limit (0.08 mag, Figure \ref{fig:Themis}), the spectral slope plateaus once the maximum slope value is reached. 

Since smaller objects likely have younger surfaces on average due to their shorter collisional disruption \citep[e.g.,][]{deelia2007}, surface refresh via seismic shaking \citep{richardson2005}, and YORP spin-up and failure \citep{graves2018} timescales, we interpret these spectral slope changes to be due to space weathering of the C-complex asteroid surfaces. We anticipate that non-catastrophic surface refresh processes are the primary methods of exposing fresh regolith and resetting the surface age. 

\begin{figure}[t]
\epsscale{0.75}
\plotone{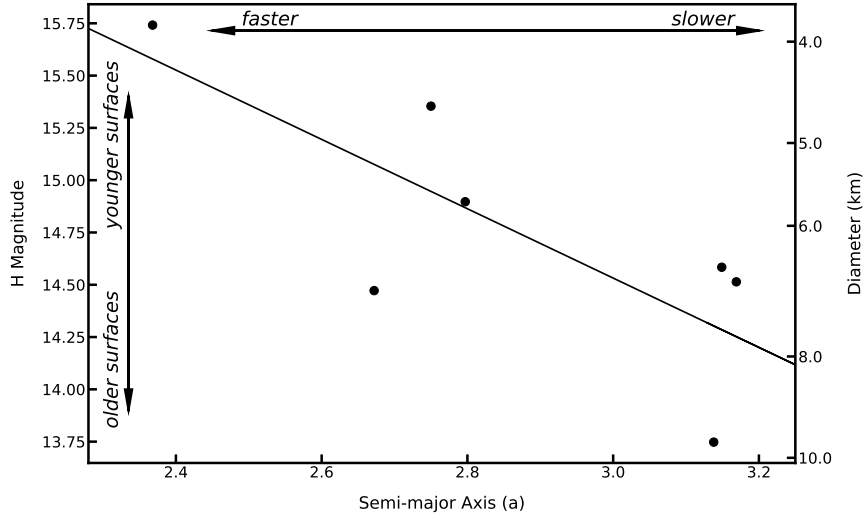}
\caption{For Hygiea-type families, the binned H magnitude of the slope minimum versus the semi-major axis of the family parent body. The right axis has the approximate diameter calculated using the average C-complex albedo, $p_V=0.06$ \citep{demeo2013}. We correlate object size to average surface age as indicated. The transition from slope blueing to slope reddening occurs at smaller sizes (and, therefore, younger average surface ages) for families that are closer to the Sun. We link smaller semi-major axis with a faster weathering process due to the increased solar wind, micrometeorite flux, and micrometeorite velocities.
   \label{fig:Vpos}}
\end{figure}

To further investigate the space weathering hypothesis for the Hygiea-type families, we examine the relationship between the location of the family in the Main Belt and the size at which the minimum spectral slope is reached. Figure \ref{fig:Vpos} shows the H magnitude of the bin with the minimum spectral slope versus the semi-major axis of the asteroid family parent body. The figure also displays approximate diameter calculated from H magnitude using the average C-complex albedo from \cite{demeo2013}. The transition from slope blueing to slope reddening occurs at smaller sizes for families that are closer to the Sun. This correlation suggests that the process responsible for this transition occurs at a younger surface age for objects with smaller semi-major axes. Given this relationship, it is likely that the increased solar wind flux, micrometeorite flux, and micrometeorite velocities at smaller heliocentric distances are a contributing factor to the timescales associated with the observed spectral slope trend.

\begin{figure}[t]
\epsscale{0.7}
\plotone{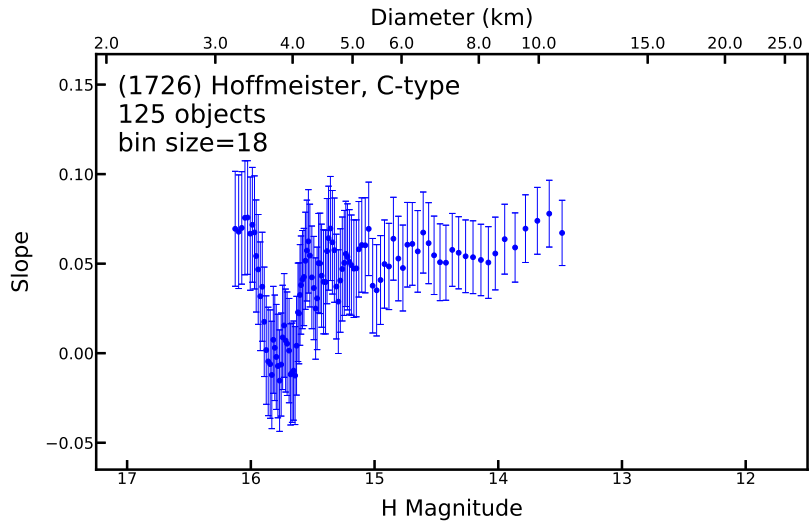}
\caption{The Hoffmeister family does not show the Hygiea-type trend with the nominal magnitude error cutoff. If we use a less selective error limit of 0.08 magnitudes, the Hoffmeister family shows the V-shaped Hygiea-type trend. \label{fig:Hoff}}
\end{figure}

Seven of the nine families investigated show the Hygiea-type spectral slope trend with the nominal SDSS magnitude error limit. Since the imposed error limit preferentially removes smaller objects, which tend to have higher observational errors, we consider if this restrictive limit has prevented the observation of the V-shape in the remaining two families. When we use a less selective error limit of 0.08 magnitudes, the Hoffmeister family clearly shows the same Hygiea-type shape (Figure \ref{fig:Hoff}). 
Therefore, we classify this family as a member of the Hygiea-type group. Despite this classification, we do not add Hoffmeister to the sample in Figure \ref{fig:Vpos} because the different error limits result in significantly different sample and bin sizes which affects the position of the average H magnitude for the blueing to reddening transition. As previously noted, using a less restrictive error limit for Themis (Figure \ref{fig:Themis}) slightly changes the spectral slope trend at the largest sizes, but the trend remains unchanged at the smallest sizes.
Additionally, if we use the trend in Figure \ref{fig:Vpos} as a guide for where we would expect to observe the Hygiea-type slope transition in Themis, the slope minimum would occur at H$\sim$14.3 which is present in the H magnitude range included in the 0.06 magnitude error limit analysis (Figure \ref{fig:panelC}). We conclude that the Themis family shows a trend that is distinctly different than the Hygiea-type slope trend present in the other families. 

We note that the Hygiea and Themis families show different spectral slope trends with respect to object size despite having several similarities that are critical to the space weathering process such as heliocentric distance, average albedo \citep{masiero2015}, and family age \citep{nesvorny2005,marzari1995}. The families are the two largest families in our sample with sample sizes of hundreds of objects and the key features of their slope trends are robust to changes in the bin size and the applied error limit. The observed difference in slope trends suggests that the spectral slopes changes in each family are likely due to crucial differences in surface composition. 

Previous work \citep[e.g.,][]{nesvorny2005SDSS} examined space weathering across the Main Belt by comparing objects from different families directly to each other. Our analysis finds that the range of spectral slopes varies greatly from family to family. Therefore, any comparison of slopes between families would see spectral slope variation that is likely related to the composition and specific spectral characteristics of the families themselves instead of being representative of the age of the asteroid surfaces. We find no clear correlations between the calculated spectral slopes or the depth of the V-shaped Hygiea-type trend for each family with their ages or heliocentric distances.

To complement our spectral slope analysis, we also attempted to investigate albedo changes with respect to size using data from the Wide-field Infrared Survey Explorer (WISE) spacecraft \citep{wright2010}. We used the \cite{nesvorny2015families} lists described in Section \ref{sec:analysis} to identify family members within the \cite{mainzer2016} Planetary Data System catalog. We restricted the WISE results to those data marked with the ``DVBI" fit code indicating that the diameter, visible geometric albedo, NEATM beaming parameter, and infrared geometric albedo were allowed to vary in the thermal fit. Unfortunately, the \cite{nesvorny2015families} objects are all optically selected and the results from the combination of this data with the NEOWISE catalog showed a clear bias against small dark objects. The lack of data was especially apparent for objects that had diameters smaller than the size at which the Hygiea-type families reach the local slope minimum and the spectral slope trend changes direction. \cite{masiero2013} used the hierarchical clustering method (HCM) on the known Main Belt asteroids detected by WISE to identify family members. When we use the \cite{masiero2013} family lists to examine potential albedo trends using strict error limits ($\sigma_{p_V}=0.05$), the sample size remaining is too small in the relevant size range even for the largest families to examine the families for trends. While we cannot examine any albedo trends in detail, we note that the standard deviations on the average C-complex family albedos \citep{masiero2013,masiero2015} are smaller than for comparably sized S-complex families (e.g., Hygiea $p_V=0.070\pm0.018$ vs. Flora $p_v=0.305\pm0.064$) so it is unlikely that there is significant variation of the albedos within each family. 

\section{Discussion} \label{sec:disc}

Analyses of S-complex objects have shown clear space weathering trends from examining the relationship between object size and spectral slope \citep{binzel2004,thomas2012}. By examining this trend within known, characterized families we reduce potential spectral variation due to compositional differences. The two types of slope trends seen in the C-complex families show trademarks of being products of the competing processes of space weathering and surface refreshing. Most importantly, the trends show a clear dependence of spectral slope on the size of the object, which we use as a proxy for average surface age, within each family. We also see evidence that the Hygiea-type trend observed in 8 of our 9 families is correlated with the distance from the Sun of the family's parent body.   
We note that this transition happens within a narrow range of estimated object diameters ($\sim$4-10km). The transition from reddening to blueing (or plateau) in the Themis family also occurs within this same size range. Therefore, the average surface age of objects of this size is important to understanding the processes responsible for these trends and both the weathering and surface refresh timescales. Additionally, the noted diameter range of the slope transitions for the 9 C-complex families includes the diameter ($\sim$5~km) associated with space weathering saturation for S-types in the near-Earth object population \citep{binzel2004} and the Koronis family \citep{thomas2011,thomas2012}.

The processes responsible for this Hygiea-type space weathering trend affect several different taxonomic classes within the C-complex. The C, Cb, and Ch classes are represented in our sample. These three classes are distinguished in the visible wavelength region by the curvature of the overall spectrum \citep{bus2000}. The C-class spectra are more concave than the other two classes and contain a weak ultraviolet absorption at the short wavelength end of the spectrum, Ch-class spectra contain the 0.7 $\mu$m absorption feature that has been connected to hydrated silicates, and the Cb-class spectra are notably flatter than the other two groups. These spectral differences are indicative of differing compositions. We observe this Hygiea-type space weathering process in a large fraction of our analyzed families of different spectral types, which suggests that the process or processes responsible for this weathering pattern is common among Main Belt C-complex objects. 

Table \ref{tab:Cfamilies} includes the calculated dynamical ages of the families in our analysis. The Veritas family is the youngest on our list by a significant margin with an age of 8.3 $\pm$ 0.5 million years \citep{nesvorny2015identification}. The spectral slope trend seen in the Veritas family is just as distinct as the trends seen in significantly older families. The presence of the Hygiea-type trend in the Veritas family suggests that the timescale necessary to reach spectral maturity is short especially considering that Veritas is located in the outer Main Belt ($a=3.17$ au). 

Seeing these spectral slope trends in various families with an extremely wide range of ages indicates that there must be processes that are actively refreshing the surfaces of the smaller family members. 
The transition in slope trend for Hygiea-type families occurs for objects with diameters of approximately 4-10 kilometers. One possibility for refreshing the surface is regolith movement from small impactors and the resulting seismic shaking of the object. The \cite{richardson2005} model shows that a 20~cm impactor could cause ``vertical launching" (greater than 1 $g_{ast}$ accelerations) on the surface of a 5~km object under restrictive seismic propagation conditions with cohesive regolith. \cite{rivkin2011} used the \cite{obrien2005} Main Belt impact recurrence times and scaled the \cite{richardson2005} results to estimate the global regolith refresh timescale to be $\sim10^6$ years for the surface of a 5~km asteroid. This refresh timescale is orders of magnitude smaller than the ages of all families presented here (except young Veritas) and is comparable to the fast weathering rates suggested by \cite{vernazza2009}.
Another process that could be responsible for resetting the surface age is YORP spin up and failure \citep[e.g.,][]{graves2018}. \cite{rubincam2000} estimated the time required to double the rotation rate of several theoretical asteroids via YORP thermal torque. One hypothetical object included in the study was ``pseudo-Deimos" which was given the same shape, moment of inertia, and albedo as actual Deimos, but placed in the Main Belt with a semi-major axis of 3 AU. ``Pseudo-Deimos" can be used to estimate the time required to double the rotation rate for asteroids in the Hygiea and Themis families given the selected semi-major axis and similar albedo \citep[Deimos $p_{V}$=0.068$\pm$0.007,][]{thomas1996}. The results presented in Figure 6 of \cite{rubincam2000} indicate that an 8~km diameter object (the approximate object size at the transition in the slope trend for both the Hygiea and Themis families) would double its rotation rate in less than 100 Myrs. This timescale is an upper limit for the objects in our C-complex families since the spin up process would happen more rapidly at smaller semi-major axes. This spin up timescale is longer than that estimated from the \cite{richardson2005} seismic shaking model, but is still significantly shorter than the ages of most of the families discussed. We expect that the asteroid surfaces are being refreshed via a combination of these two mechanisms. 

One additional possibility to explain the change in spectral slope for the largest asteroids within the families is variation in grain size. \cite{vernazza2016} concluded that observed variation in spectral slopes for large Ch and Cgh-type asteroids is due to the anti-correlation between regolith grain size and the object diameter and, therefore, surface gravity. The largest asteroids (D$>$100~km) in their sample have redder spectral slopes and are more consistent with a fine-grained CM chondrite-like regolith compared to the more neutral spectral slopes of the smaller objects (D$<$60~km) which were linked to a coarse-grained CM chondrite-like regolith. 

Studies of carbonaceous chondrite meteorites have shown that laboratory space weathering simulation experiments on different meteorite types have resulted in divergent observed spectral and albedo changes. \cite{lantz2017} and \cite{lantz2018} clearly summarize the observed characteristics. Irradiation of CV and CO chondrites has resulted in spectra with redder slopes that are more concave and have lower albedos, while CI and CM meteorite spectra become bluer and more convex while their albedos get higher \citep[e.g.,][]{brunetto2014,brunetto2015,lantz2015,matsuoka2015}. Similarly, experiments with Tagish Lake show a flattening and brightening of the spectrum with increased irradiation \citep{vernazza2013}. In all of these laboratory investigations of space weathering processes with meteorites, increased irradiation causes a consistent increase or a decrease in spectral slope. 
The vast majority of past work involving both laboratory experiments and telescopic observations have concluded that the trend in spectral slope only moves in one direction. The results from the families analyzed in this study directly contradict those steady space weathering trends. Our analysis shows spectral slope trends that change direction after some time period. 

Irradiation experiments by \cite{kaluna2017} of aqueously altered minerals have demonstrated that a change in direction of the spectral slope evolution is possible. The spectral slopes of their Fe-rich assemblage (consisting of cronstedtite, pyrite, and siderite) initially reddened and then became bluer with increased irradiation. To understand the physical processes responsible for their spectral trends, \cite{kaluna2017} used scanning and transmission electron microscopy to examine their samples and radiative transfer modeling to model the effects of the irradiation products. For the Fe-rich assemblage, they observed nanophase iron and micron sized carbon-rich particles and estimated the optical effects of these components from the model. \cite{kaluna2017} conclude that the nanophase iron particles were responsible for the spectral reddening and the micron sized carbon particles caused the spectral blueing. They explain the slope transition as the carbon particles likely dominating the optical properties once a critical amount is present on the surface. 
This spectral analysis was performed on data with visible and near-infrared wavelengths, but the slope variations shown in their Figure 3 indicate that the observed slope trend would also be present in visible wavelength data, such as that provided by SDSS. 

Since we also observed a change in direction of the spectral slope, we are likely observing the results of two competing surface process similar to what was observed by \cite{kaluna2017}. However, the Hygeia-type spectral slope trends are the opposite of those observed in the \cite{kaluna2017} experiments: we observe blueing of the spectral slope followed by slope reddening. It is possible that our trends are also the result of the competing optical effects of nanophase iron and micron sized carbon-rich particles under different conditions. Many of the compositional components expected in the C-complex would have sufficient iron and carbon to create these irradiation products. Additional laboratory experiments may help explain the physical processes behind this Hygiea-type space weathering spectral slope trend.

The Themis family is the only family in our C-complex sample that does not show a Hygiea-type trend. It is not surprising that Themis is different given its unique composition and evolution. Water ice has been detected on the surface of Themis \citep{rivkin2010,campins2010} and past work suggests that the parent body experienced some thermal evolution \citep{castillo2010,marsset2016}. A partially differentiated Themis parent body would result in a family that contained various compositions and, therefore, spectral properties. The unique Themis spectral slope trend could be due to space weathering of the thermally evolved asteroids. 
It is also possible that the observed Themis spectral slope trend is indicative of variation in composition, but there is no known reason why composition would vary with object size within the family.
The change in slope direction for Themis is not as distinct as in the Hygiea-type trends and is absent in our analysis with higher photometric error limits. Our data is also mostly consistent with \cite{kaluna2016}, who conclude that Themis family visible wavelength spectra become redder due to space weathering.

There are numerous C-complex families that are not included in this analysis because there were not an adequate number of family members in the SDSS MOC that met our strict magnitude error limits. We anticipate that future large photometric catalogs, such as the one that will be generated by the Rubin Observatory (formerly LSST), will enable the study of space weathering trends in C-complex families with greater detail. The expected single exposure limiting magnitudes (5$\sigma$) for the Rubin Observatory are 23.4, 24.8, 24.3, 23.9, and 23.3 in any of the $u$, $g$, $r$, $i$, and $z$ filters, respectively\footnote{https://smtn-002.lsst.io/}. These exposure depths are all at least 2 magnitudes fainter than the 95\% completeness limit for SDSS, which suggests the upcoming survey will be able to observe objects with diameters $\sim$2.5 smaller than SDSS. In addition to adding new families to this type of study, we anticipate spectrophotometry for objects down to $\sim$1~km in diameter for several of the C-complex families included in our analysis. 

\section{Conclusions} \label{sec:conc}

Our analysis of space weathering trends in C-complex families shows two distinct spectral slope trends: Hygiea-type and Themis-type. Eight of the nine families studied show the Hygiea-type space weathering trend. The processes responsible for the distinct changes in spectral slope affect several different taxonomic classes within the C-complex and appear to act quickly to alter the spectral slopes of the family members. The Hygiea-type trend appears to be correlated with the distance from the Sun of the family indicating that the observed trends are related to space weathering of the objects. The Themis family is the only family to not show the V-shaped Hygiea-type slope trend. We encourage further laboratory experiments to explain the processes driving these two space weathering trends.

\acknowledgments

Funding for the creation and distribution of the SDSS Archive has been provided by the Alfred P. Sloan Foundation, the Participating Institutions, the National Aeronautics and Space Administration, the National Science Foundation, the U.S. Department of Energy, the Japanese Monbukagakusho, and the Max Planck Society. The SDSS Web site is http://www.sdss.org/.

The SDSS is managed by the Astrophysical Research Consortium (ARC) for the Participating Institutions. The Participating Institutions are The University of Chicago, Fermilab, the Institute for Advanced Study, the Japan Participation Group, The Johns Hopkins University, the Korean Scientist Group, Los Alamos National Laboratory, the Max-Planck-Institute for Astronomy (MPIA), the Max-Planck-Institute for Astrophysics (MPA), New Mexico State University, University of Pittsburgh, University of Portsmouth, Princeton University, the United States Naval Observatory, and the University of Washington.

Some of this work was carried out while TL was
supported by the NSF Research Experiences for Undergraduates program
at NAU under NSF award AST-1004107.

\appendix
\section{Completeness Limits for Families} \label{sec:append}

\begin{figure}[t]
\epsscale{1.2}
\plotone{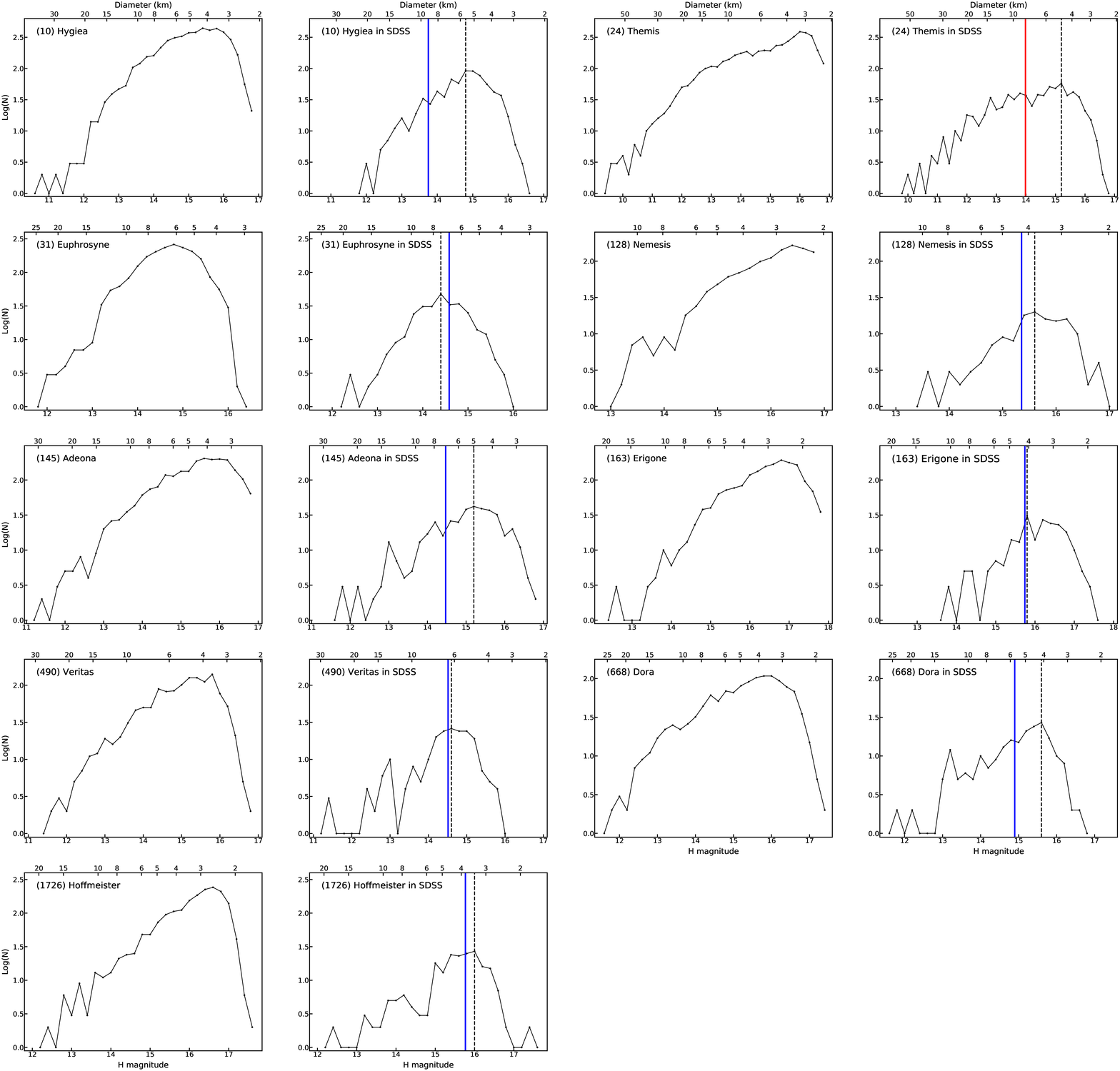}
\caption{Size frequency distributions (SFDs) for each family are used to determine completeness. The figure panel contains two SFDs for each family: (1) from the \cite{nesvorny2015families} list and (2) using the list of family objects identified in the SDSS MOC. We use the SFD from the SDSS MOC to determine if the family is complete. The completeness limit is estimated from the maximum value of the SFD and a family is determined to be complete if the completeness limit (dashed line) is at a larger H magnitude (smaller diameter) than H magnitude of the change in the spectral slope trend (blue for Hygiea-type, red for Themis). \label{fig:sfd_append}}
\end{figure}

We examine the size frequency distribution (SFD) of each of the 9 C-Complex families to determine if our observed spectral trends are the result of observational biases. 
We present the size frequency distribution of each family in two different methods (Figure \ref{fig:sfd_append}): (1) using the \citet{nesvorny2015families} family list with H magnitudes from the MPC and (2) using the list of family objects identified in the SDSS MOC with H magnitudes from ASTORB. For the asteroids identified in the SDSS MOC, we removed those objects with observed magnitudes beyond the limiting magnitudes for 95\% completeness \citep[greater than 22.0, 22.2, 22.2, 21.3, and 20.5 in any of the $u$, $g$, $r$, $i$, and $z$ filters,][]{ivezic2001}. We used the provided catalog H magnitudes for each of these datasets. The H magnitudes are similar, but are not identical for many objects. Each size frequency distribution uses H magnitude bins of 0.2 magnitudes, which mitigates many of the H magnitude differences in the two datasets.
Our family analysis removes additional photometry from the SDSS MOC, but we choose not to remove additional data from this SFD investigation since this subset of the data best represents the completeness of the sample.   

We use the size frequency distribution from the families within the SDSS MOC to determine if a family is complete. 
We classify a family as complete if the completeness limit for the family is at an H magnitude larger (diameter smaller) than any changes in our spectral slope trends. We estimate the completeness limit for each family as the maximum value of the SDSS MOC family SFD. This process slightly overestimates (with respect to H magnitude) the true completeness limit for the families. Figure \ref{fig:sfd_append} shows the determined completeness limit as a dashed line and the H magnitude at which the spectral slope changes as a blue (Hygiea-type) or red (Themis-type) line. We also classify a family as incomplete if the spectral slope change is within $\sim$0.2 (the SFD bin size) magnitudes of the completeness limit (Erigone, Veritas, Hoffmeister).   
We use the \citet{nesvorny2015families} SFDs for each family to compare the SDSS MOC results to that of the entire family. 

\begin{figure}[h]
\epsscale{1.2}
\plotone{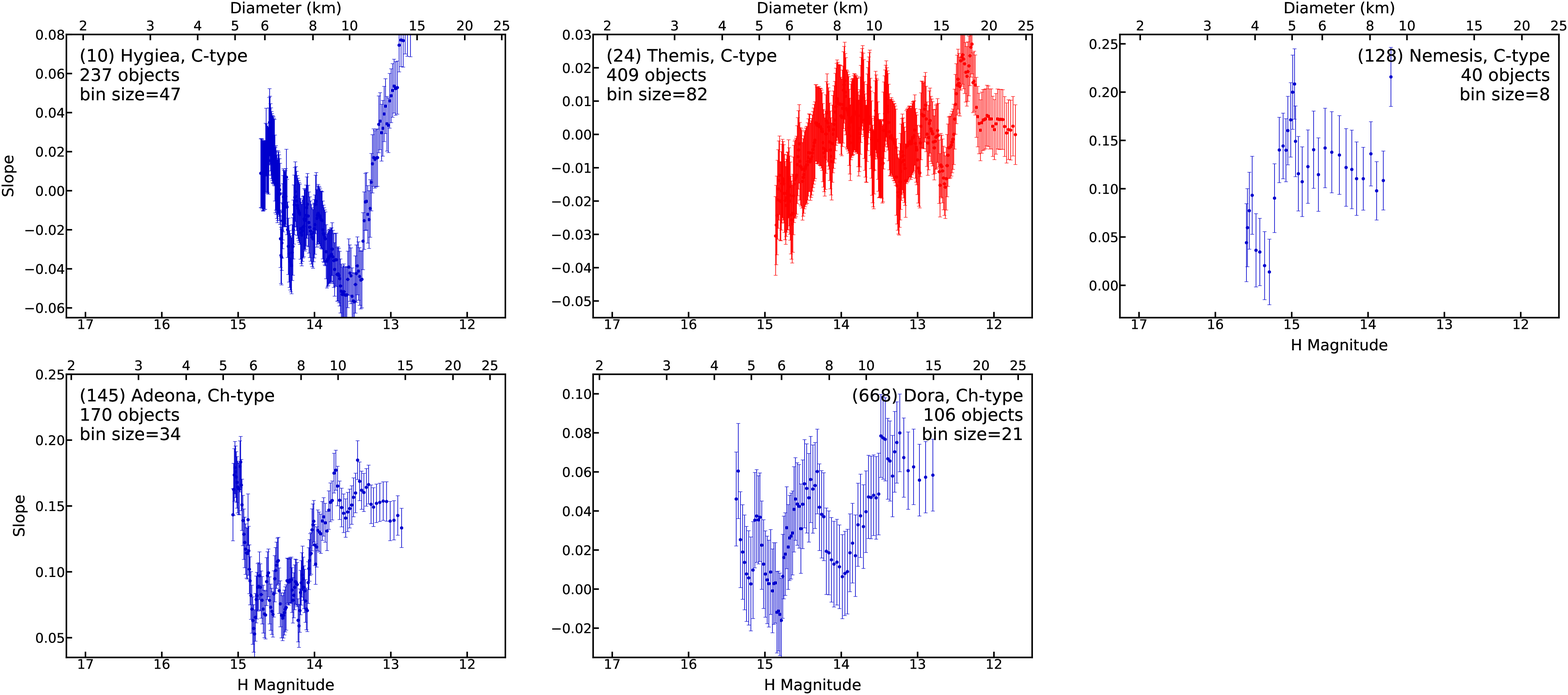}
\caption{Spectrophotometric slope versus H magnitude and approximate object diameter for the five complete C-complex asteroid families using only objects with H magnitudes below the defined completeness limit for the family. Each figure includes the family name, taxonomic type, number of objects in the sample, and the bin size for the running box mean calculation. The range of H magnitude is held constant for each family and is shown along the bottom axis for each figure. The top axis shows the approximate diameters for the corresponding H magnitudes. For comparison, this figure uses the same range of slope values for each family as seen in Figure \ref{fig:panelC}. The Hygiea-type families (blue) and Themis family (red) show a similar spectral trends to those in Figure \ref{fig:panelC}. \label{fig:complete_append}}
\end{figure}

The slope trends presented in Figure \ref{fig:panelC} for the complete families includes slopes for objects beyond the defined completeness limit. Due to the running box mean calculation, slopes for objects beyond the family's completeness limit will be included in calculated average slopes near and under that limit. We verify the robustness of our slope trends to potential observational biases introduced through extending beyond the completeness limit by repeating our analysis for the 5 complete families using only family members with H magnitudes smaller than the completeness limit. We find that the recalculated slope trends of the complete families are consistent with the trends presented this work and that the Hygiea-type and Themis-type trends remain distinct from each other (Figure \ref{fig:complete_append}). 

\bibliography{Thomas_SDSS}






\end{document}